\documentclass{sig-alternate}
	
\usepackage[tight,footnotesize]{subfigure}
\usepackage{graphicx,epstopdf} 
\usepackage{cite}

\begin{document}

\title{Up and Away: A Cheap UAV Cyber-Physical Testbed\\(Work in Progress)}

\numberofauthors{1}
\author{
\alignauthor  {Ahmed Saeed$^\dag$, Azin Neishaboori$^{\dag\ddag}$, Amr Mohamed$^\dag$, Khaled A. Harras$^\ddag$ }\\
\affaddr{$^\dag$Department of Computer Science and Engineering, College of Engineering, Qatar University}\\
\affaddr{$^\ddag$School of Computer Science, Carnegie Mellon University Qatar}\\
       \email{\normalsize ahmed.saeed@qu.edu.qa, azin.neishaboori@gmail.com, amrm@qu.edu.qa, kharras@cs.cmu.edu}}


\maketitle
\begin{abstract}
Cyber-Physical Systems (CPS) have the promise of presenting the next evolution in computing with potential applications that include aerospace, transportation, robotics, and various automation systems. These applications motivate advances in the different sub-fields of CPS (e.g. mobile computing and communication, control, and vision). However, deploying and testing complete CPSs is known to be a complex and expensive task. In this paper, we present the design, implementation, and evaluation of \emph{Up and Away (UnA)}: a testbed for Cyber-Physical Systems that use UAVs as their physical component. \emph{UnA} aims at abstracting the control of physical components of the system to reduce the complexity of UAV oriented Cyber-Physical Systems experiments. In addition, \emph{UnA} provides an API to allow for converting CPS simulations into physical experiments using a few simple steps. We present a case study bringing a mobile-camera-based surveillance system simulation to life using \emph{UnA}.
\end{abstract}
%
\keywords{UAV Testbed, Cyber Physical Experiments}

\section{Introduction}

Cyber-Physical Systems (CPS) have the promise of presenting the next evolution in computing by bridging the gap between the virtual world and the physical world. CPS potential applications include aerospace, transportation, and factory automation systems \cite{rajkumar2010cyber}. Advancements in cyber-physical systems research is supported by development in other research fields including mobile computing, embedded systems, computer vision, control, and communication. While this development ultimately contributes to advancing different components of cyber-physical systems, most of the work is evaluated either through simulations or through experiments that focus on the specific sub-problem. Furthermore, the complex and multidisciplinary nature of full scale experiments of cyber-physical systems makes the task even harder especially when it comes to testing individual "cyber" or "physical" components. From a different perspective, current generic CPS testbed \cite{fok2011pharos} are typically complex and incur large deployment cost.


In this paper, we present Up and Away (UnA), a testbed that allows for low-cost, rapid development of cyber-physical systems experiments. \emph{UnA} abstracts the "physical" components of the system to allow for experiments that focus on "cyber" components with real physical interactions. We chose UAVs as the physical component of the system due to their deployment flexibility and maneuverability. Furthermore, UAVs are suitabile for many application (e.g. surveillance, transportation, and communications) allowing for cyber-physical experimentation with realistic objectives.

Several research groups have recently developed quadcoptor testbeds \cite{jung2005design,lupashin2010simple,
valenti2007indoor,michael2010grasp}. The GRASP testbed \cite{michael2010grasp} uses off-the-shelf high-end Ascending Technologies Hummingbird quadcopters to demonstrate multirobot control algorithms. The ETH Flying Machine Arena \cite{lupashin2010simple} uses modified Hummingbird quadcopters to demonstrate several acrobatic and athletics control maneuvers. A testbed was developed by University of Colorado, Boulder for testing ad-hoc networking scenarios \cite{brown2004test}.  The testbed uses UAVs which were deployed in a 7 $km^2$ area to test different operation conditions of the DSR protocol. The work in \cite{jung2005design} shows the development of a low-cost UAV testbed using Goldberg Decathlon ARF model airplane. The work aims at showing the design and development of hardware and software components to produce low-cost UAVs. 

Unlike other UAV-based testbeds, \emph{UnA}'s architecture is more concerned with abstracting the control of UAVs rather than improving it. Furthermore, \emph{UnA} presents an architecture that provides default components and facilitates replacing any of them to test its impact on the overall performance of the system. To that end, AR Drones were selected as the UAVs to be used. We preferred using quadcopters to fixed-winged airplanes due to their maneuverability especially in highly constrained spaces (e.g. indoors) including their ability to hover. Moreover, AR Drones are extremely cheap\footnote{An AR Drone 2.0 costs \$300 while an AscTec Hummingbird costs around \$4500 .} UAVs that come with a large array of sensors and a widely supported open source API to control the drone and obtain sensory information for its sensors (the specifications of the AR Drone are better described in Section~\ref{sec:arch}).








\section{UnA Architecture}
\label{sec:arch}
In this section, we introduce the design and implementation of UnA. We start with our design goals followed by the architecture description. The, We describe the details of how \emph{UnA} handles targets and UAVs localization, UAVs control, CPS processing and communication.

\subsection{Design Goals}
The development of \emph{UnA} is motivated by the need for a low-cost, easy to deploy experiments for CPS. Thus, \emph{UnA}'s architecture is designed with the following goals in mind:

\vspace{2mm}

\noindent 1. \textit{Multidisciplinary nature of CPS experiments:} The development of CPS experiments requires awareness of several fields including hardware design, control systems, software systems development \cite{rajkumar2010cyber}. This poses significant challenges for developing realistic experiments for such systems. \emph{UnA} was designed with this unique feature of CPS in mind and therefore the \emph{UnA} modules are extensible to support development of any of the CPS's components.

\vspace{1.5mm}

\noindent 2. \textit{Support of large scale experiments:} CPS experiments with a large number of physical nodes requires the testbed to support different communication models between the nodes. Furthermore, the processing power required to optimize the overall state of the system can grow exponentially with the number of nodes. The \emph{UnA} architecture supports Ad-Hoc communication using AODV for routing to facilitate communication between the different UAVs. Furthermore, it supports off-loading processing tasks required by the UAVs to central or multiple computing nodes.

\vspace{1.5mm}

\noindent 3. \textit{Seamless control of the UAVs:}
\emph{UnA} aims at reducing the complexities in CPS experiments design incurred by the automatically operated mechanical components of the system by abstracting the control algorithms and hardware design which allows for more focus on the "cyber" problems. Furthermore, autonomous control of off-the-shelf UAVs is a difficult task \cite{michael2010grasp,lupashin2010simple}. Thus, \emph{UnA} relies on the easy to use AR Drone API and extends it to use way-point navigation instead of controlling the rotation (i.e. roll, yaw and gaz) of the UAV.


\subsection{Overview}

\begin{figure}
 \centering
   \includegraphics[width=0.54\textwidth]{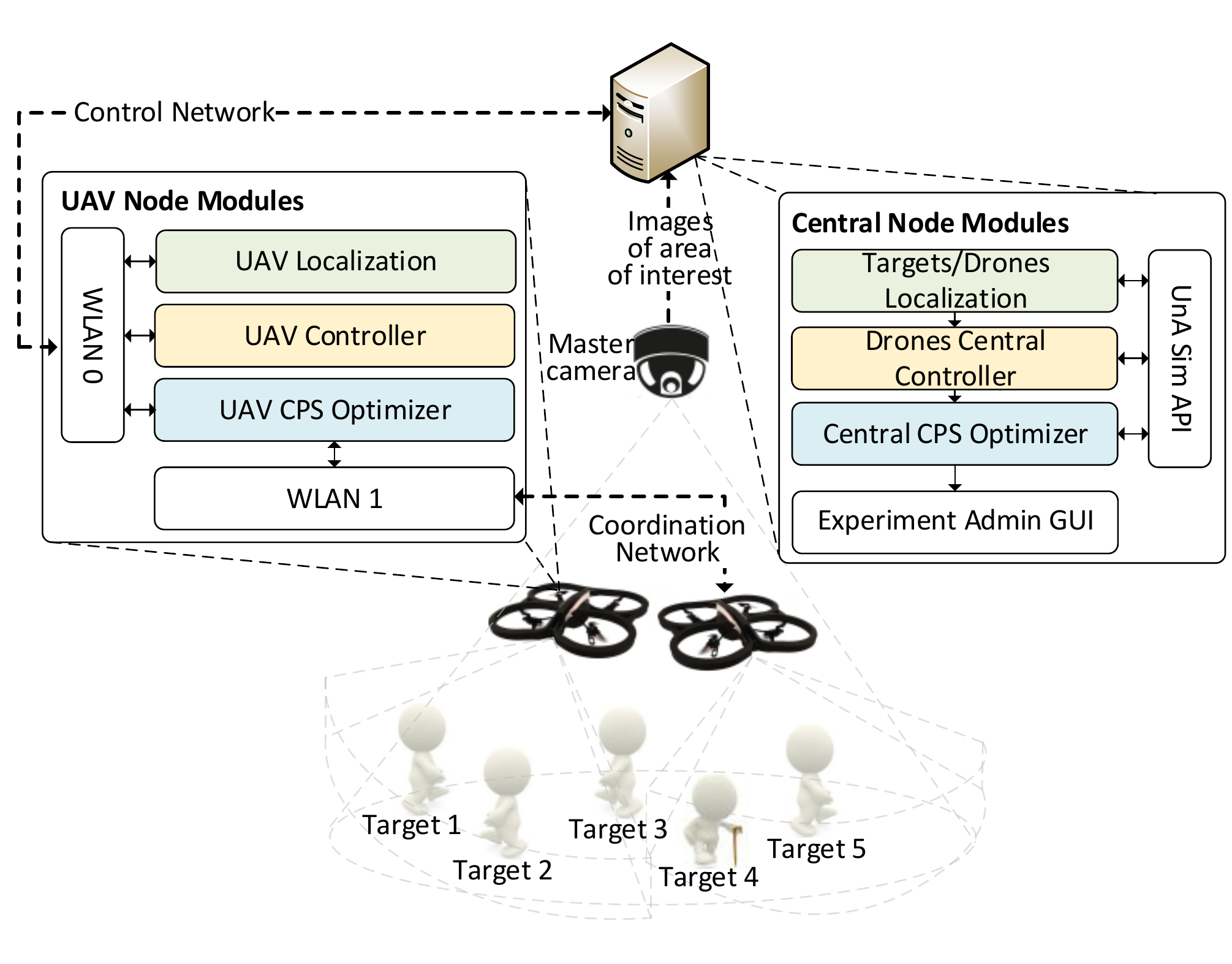}\caption{UnA Architecture}
   \label{fig:arch}
\end{figure}

Figure~\ref{fig:arch} shows a summary of the \emph{UnA} architecture. \emph{UnA} allows the UAVs that can communicate with each other, process sensory information locally and act accordingly (i.e. control itself). It also provides a base node that monitors the behavior of all UAVs. Moreover the central node can handle processing intensive tasks that UAVs off-load to it.

The \textit{UAV Node Modules} are responsible for obtaining and processing information about the drone and environment's state, calculating the drone's corresponding objectives accordingly, and controlling the drone's motion to achieve these objectives. \emph{UnA} is built on top of the Parrot AR Drone 2.0 \cite{krajnik2011ar}. The AR Drone is a quadrotor helicopter that are electerically powered with two camera: a front 720p camera with $93^{\circ}$ lens and a vertical QVGA camera with $64^{\circ}$ lens. These drones are controlled through an external computer through WiFi and are equipped with an on board ARM Cortex A8 processor with 1 Gbit RAM that runs a Linux 2.6.32 Busybox. The on-board computing machine is responsible for collecting and reporting the state of the drone to an external computer that controls the drone. The AR Drones are relatively cheap, easy to program and comes with an SDK that provides assisted maneuvers which satisfy most of our design goals.

The \textit{Central Node Modules} are responsible for monitoring the state of each drone, manually controlling the drone, and providing part of the environment state information to the drones. Due to the limited processing capacity of current off-the-shelf UAVs, UnA's architecture enables off-loading some of the processing tasks from the UAVs to the central node to enable real time responses to changes in the environment. The \textit{Experiment Admin GUI} displays the state information of the different UAVs. It also supports the manual control of the UAVs for emergency cases. The main purpose of UnA's architecture is to support the rapid development of CPS experiments that use UAVs as their Physical part. While all UnA can be extended and customized for different CPS applications, we assume that simulation code was already developed by the experiment administrator. Thus, the central node supports communication between its different modules with other applications through sockets. The \textit{UnA Sim API} provides the appropriate wrappers to allow the simulation code to interface with UnA to bring the simulations to life by off-loading all the \textit{UAV Node Modules} functionalities to the simulation engine through the central node.

For the rest of the section, we explain the different layers of the \textit{UnA} architecture according to the color coding in Figure~\ref{fig:arch}.

\subsection{UAV and Target Localization}

UnA provides two different approaches for obtaining location information of the UAVs: 1) Parrot Flight Recorder GPS module and 2) using a master camera mounted at a high point to track the location of the drone. While the first approach is straight forward, it doesn't work within controlled indoor environments. Thus,  computer vision is used for tracking the UAVs in \emph{UnA}.

Multiple UAV tracking using computer vision with multiple cameras is a tricky task to implement  \cite{oh2011indoor}. For that purpose, we propose a simpler approach that uses a single camera, mounted at a high point in the center of the area of interest, to track the drones by tagging each drone with a distinct color tag\footnote{While multiple color tags are used in \cite{oh2011indoor} to track the drone's location and orientation, our proposed approach uses a single tag to track the location and the internal campus of the drone to detect the drone's orientation.}. Tracking objects of a certain color could be easily implemented using OpenCV \cite{bradski2008learning}. This approach is further extended to track other targets as well.

While earlier work \cite{lupashin2010simple, michael2010grasp} relies on Motion Capturing Systems (MCS) for tracking UAVs. We chose the single camera approach as a much cheaper alternative that provides reasonable accuracy (Section~\ref{sec:case})\footnote{ While PTZ cameras can cost several thousand dollars, MCS costs up to tens of thousands of dollars.}.

\subsection{UAV Control}
\label{sec:control}
\textit{UAV Controller} and \textit{Drones Central Controller} are responsible for moving the drone to a specific X, Y coordinates with a certain orientation within the area of interest.  While quadcopters can move in any direction with any orientation, we use a simple controlling method: at the beginning of each command the orientation of the drone is set to $90^{\circ} w.r.t.$ the x-y plan of the area of interest. The change in orientation uses the compass for feedback.  The UAV then moves the drone along the x-axis to its specific coordinate then along the y-axis and finally it's rotated to the specified orientation. This control algorithm obtains feedback from the UAV localization module and the UAV internal compass in order to actuate any errors in placing the drone due to its inertia.

The AR Drone's motion is controlled by the AT commands protocol. The AT commands are text based commands that are translated by the AR Drone's firmware to control the drone's roll, pitch, gaz, and yaw to specify the direction, speed, and duration of the drone's motion. The controller converts the newly obtained objectives and the current location information to the corresponding set of AT commands. This conversion could be made locally on the drone (i.e. autopilot mode) or on the central node (i.e. remote control mode).

\subsection{CPS Optimization}

The \textit{UAV CPS Optimizer} module is responsible for updating the objective (i.e. location and orientation) of the drone after obtaining updates on the state of the drone and the environment from the drone's sensor, other drones and the central node. This module has three modes of operation: 1) distributed mode, 2) central mode and 3) emulation mode. In the distributed mode, a distributed objective function is used where each UAV calculates its own new objectives after each update which is suitable for lightweight distributed algorithms. In the central mode, the \textit{UAV CPS Optimizer} reports the updates to the \textit{Central CPS Optimizer} which calculates the new objectives based updates from all UAVs. Finally, in the emulation mode, the UAVs implement a distributed objective function but off-load its calculations to the central node to meet real-time responsiveness requirements.

\subsection{Control and Coordination Networks}

One of our design choices is to separate the intercommunication between the UAVs and the communication between each UAV and the central node to two different networks. This design choice was motivated by two reasons: 1) While communication between UAVs can be established in ad-hoc, unreliable fashion, communication between the UAVs and the central nodes should be using an infrastructure that is connected to the UAVs at all times, 2) AR Drone 2.0 \cite{krajnik2011ar} provides an SDK for monitoring and controlling the UAVs  that requires continuous communication between the UAVs and a central node. We rely on that SDK to facilitate the different tasks of the \textit{Central Node Modules}.

For the implementation of the \textit{Control Network}, we use the built in WiFi card but changed its configuration to work as a client instead of as an access point which is its default configuration. This way all nodes can connect to the central node through a WiFi infrastructure.  As for the \textit{Coordination Network}, we plug a WiFi dongle into the drone, compile the dongle's driver to work on the ARM-based Busybox, and install the driver on the drone. The dongle is then configured to work in Ad-Hoc mode and join the drones SSID as soon as its visible.




%
%
%
%
%
\section{Case Study}
\label{sec:case}

In this section, we demonstrate the steps of bringing a sample CPS simulation to life using \emph{UnA}, namely a mobile-camera-based surveillance system. The system's objective is to maximize the number of targets covered by a set of mobile cameras. The proposed algorithm is a central algorithm whose  inputs are the location of all targets and the mobile cameras, and its output is a set of location and orientation directives for all mobile cameras. For the rest of the section, we will show the different steps of implementing the proposed CPS using \emph{UnA}.

\subsection{UAVs and Targets Localization}

To localize the different moving objects in the area of interest (i.e. targets and UAVs), each drone is tagged with a distinct color. On the other hand, because we do not care about the identity of the targets, all targets are tagged with the same color. An Axis 213 PTZ Network Camera is mounted at 5.2 meters to locate the different objects. The central node queries the camera for images at a rate of 20 Hz and uses OpenCV to search each image for all tags by iteratively performing the following steps for each color tag:
\begin{enumerate}
\item Filter the image based on the color range of the tag (\verb|inRange()|). The tag's color range is extracted by initial calibration.\vspace{-0.08in}
\item Filter noisy parts of the image that fall within the same color range as the tag by eroding areas smaller than $3\times3$ pixels (\verb|erode()|). By this step, the only remaining white area within the image is the tag we are trying to localize. \vspace{-0.08in}
\item Detect contours using \cite{suzuki1985topological} to locate the tag within the image (\verb|findContours()|). Pixel coordinates can then be scaled to actual x,y coordinates.
\end{enumerate}

\subsection{CPS Implementation}

Our initial simulations results for the coverage algorithm were implemented on Matlab to benchmark the performance of the coverage algorithm using a different set of parameters (e.g. number of targets, number of mobile cameras and targets distribution). The Matlab code was used with a minor modification of obtainning its input from the localization module and sending its output to the UAV controller through sockets.

\subsection{UAV Control}

To simplify the case study, all UAV controlling operations are off-loaded to the central node. The central node uses AR Drone API v2.1. to control the motion of the UAVs. The SDK provides the following command \verb|ardrone_tool_set_ui| \verb|_pad_start| for take off and landing and \verb|ardrone_tool_set| \verb|_progressive_cmd| for flying and hovering by controlling the UAVs yaw, gaz, roll and pitch. The controller is implemented following the motion methodology presented in Section~\ref{sec:control}.

\subsection{Experimental Scenario}

\begin{figure}
\centering
\begin{subfigure}
  \centering
  \includegraphics[width=.49\linewidth]{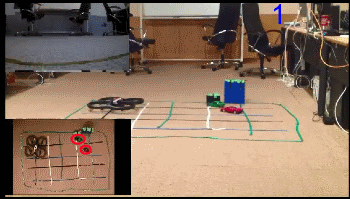}
\end{subfigure}%
\begin{subfigure}
  \centering
  \includegraphics[width=.49\linewidth]{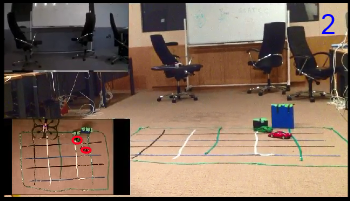}
\end{subfigure}
\begin{subfigure}
  \centering
  \includegraphics[width=.49\linewidth]{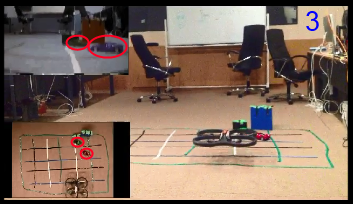}
\end{subfigure}%
\begin{subfigure}
  \centering
  \includegraphics[width=.495\linewidth]{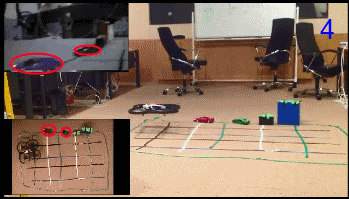}
\end{subfigure}
\caption{The top left corner is the view from the UAV while the bottom left corner is the view from the master camera. Targets are circled in red when they appear in the view of either cameras.}
\label{fig:sample_run}
\end{figure}

The area of deployment was 1.25 by 2.1 meters (4.1 by 6.8 feet). Figure~\ref{fig:sample_run} show snapshots of an \emph{UnA}. The scenario of this experiment is to first have the two targets (circled in red) positioned in a certain configuration and have the UAV cover them. Then, move the targets and make the UAV reposition itself to cover the targets.

The first two snapshots show the UAV moving itself to cover the targets in their initial configuration. The third snapshot shows the UAV covering both targets (i.e. both targets are within the view of the drone's camera). The fourth snapshot shows the UAV after moving to cover the drones on their second configuration.




\section{Limitations and Future Work}
\label{sec:conc}
\begin{figure}
 \centering
   \includegraphics[width=0.3\textwidth]{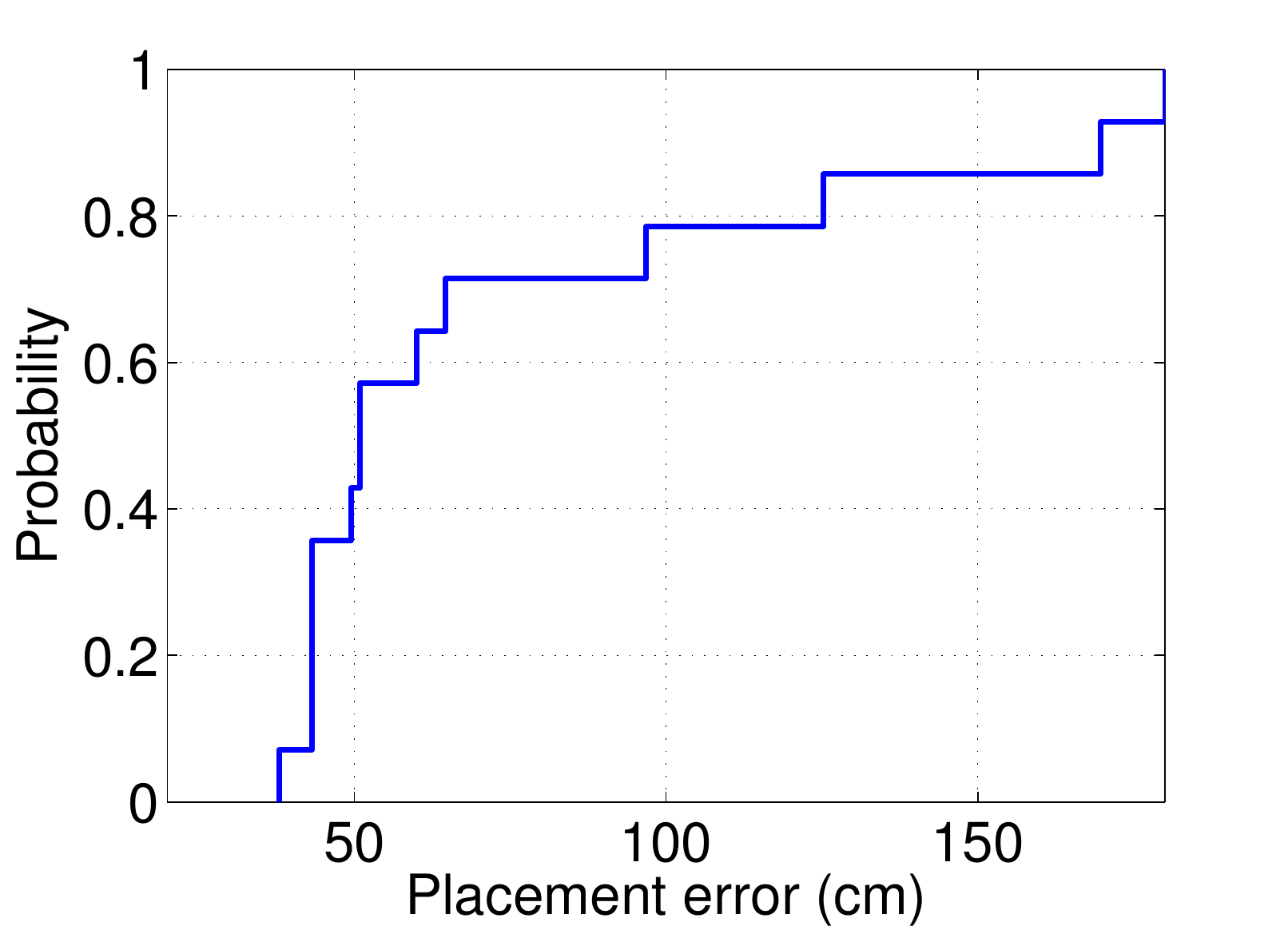}\caption{CDF of distance error in drone placement.}
   \label{fig:error}
\end{figure}

\emph{UnA} reduces the cost and overhead of CPS experimental deployments, however these advantages does not come without some limitations. One of the major limitations of \emph{UnA} is the accuracy of the UAV placement. Errors in UAV control and maneuvering inherently appear due to inaccuracies in computer vision based localization system and assisted maneuvers provided by the AR Drone's firmware.
We benchmarked the \emph{UnA}'s placement error by placing the drone arbitrarily and then moving it to a certain location over 14 times. Figure~\ref{fig:error} shows the CDF of the error in placing the drone (i.e. the distance between the drone's actual location and its intended location).
The effects of these placement inaccuracies are significantly increased due to the small size of the deployment area. Nonetheless, the current state of \emph{UnA} could be useful for many applications.

Our future plans for \emph{UnA} include larger deployments with two or three drones. We also plan to deploy Click Modular router on the UAVs to allow for a more extensive evaluation of communication protocols between the UAVs. Furthermore, we plan to extend the "physical" components abstraction to allow the seamless integration of several other mechnical systems (e.g. ground vehicles). We also plan to explore other approaches for UAV tracking (e.g. RFID tracking).

\bibliographystyle{abbrv}
{\small
\bibliography{UnA_ref}}
\end{document}